\documentclass{ifacconf}

\usepackage{graphicx}      
\usepackage{natbib}        
\usepackage{amsmath,amssymb}
\usepackage{xcolor}
\usepackage{algorithm}
\usepackage{algpseudocode}
\usepackage{subcaption}
\usepackage{comment}
\renewcommand{\tilde}{\widetilde}
\newcommand{\mD}{\mathcal{D}}
\newcommand{\mG}{\mathcal{G}}
\newcommand{\mN}{\mathcal{N}}
\newcommand{\mO}{\mathcal{O}}
\newcommand{\mT}{\mathcal{T}}
\newcommand{\R}{\mathbb{R}}
\newcommand{\Exp}{\mathbb{E}}
\newcommand{\br}{\mathbf{r}}
\newcommand{\bu}{\mathbf{u}}
\newcommand{\bx}{\mathbf{x}}
\newcommand{\bSigma}{\mathbf{\Sigma}}
\newcommand{\tp}{\mathsf{T}}
\newcommand{\dd}{\mathrm{d}}

\newcommand{\norm}[1]{\left\Vert #1 \right\Vert}
\newcommand{\abs}[1]{\left\vert #1 \right\vert}
\newcommand{\res}{\operatorname{res}}
\newcommand{\tr}{\operatorname{tr}}
\newcommand{\poly}{\operatorname{Poly}}

\newcommand{\sgn}{\operatorname{sgn}}
\algnewcommand{\LineComment}[1]{\State \(\triangleright\) #1}
\begin{document}
\begin{frontmatter}

\title{Stackelberg Meta-Learning Based Control for Guided Cooperative LQG Systems \\}

\thanks[footnoteinfo]{© 2023 the authors. This work has been accepted to IFAC for publication under a Creative Commons Licence CC-BY-NC-ND.}

\author[First]{Yuhan Zhao}
\author[First]{Quanyan Zhu}

\address[First]{New York University, Brooklyn, NY 11201 USA \\ (e-mail: \{yhzhao,qz494\}@nyu.edu)}

\begin{abstract}                
Guided cooperation allows intelligent agents with heterogeneous capabilities to work together by following a leader-follower type of interaction. However, the associated control problem becomes challenging when the leader agent does not have complete information about follower agents. There is a need for learning and adaptation of cooperation plans.
To this end, we develop a meta-learning-based Stackelberg game-theoretic framework to address the challenges in the guided cooperative control for linear systems. We first formulate the guided cooperation between agents as a dynamic Stackelberg game and use the feedback Stackelberg equilibrium as the agent-wise cooperation strategy. We further leverage meta-learning to address the incomplete information of follower agents, where the leader agent learns a meta-response model from a prescribed set of followers offline and adapts to a new coming cooperation task with a small amount of learning data.
We use a case study in robot teaming to corroborate the effectiveness of our framework. Comparison with other learning approaches also shows that our learned cooperation strategy provides better transferability for different cooperation tasks.

\end{abstract}


\end{frontmatter}

\section{Introduction} \label{sec:intro}

Cooperative control aims to address the collaboration between multiple intelligent agents and has become indispensable in modern system design to accomplish complex tasks (e.g., \cite{wang2017cooperative,chen2019control,liao2021model,papalia2022prioritized}).
Particularly, guided cooperation is gaining increasing attention and popularity as we witness the rapid advances in the development of Artificial Intelligence (AI) aided technology in control systems.
Guided cooperation allows intelligent agents with heterogeneous capabilities to work together and has a leader-follower or mentor-apprentice structure of interactions. A more resourceful agent (leader) can utilize its resources (e.g., sensing and computational resources) to provide strategic guidance to a less resourceful agent (follower) so that all agents can fully utilize their advantages to achieve the cooperation task objective.
Guided cooperation is also broadly used in many applications such as multi-agent teaming (e.g., \cite{choi2017automated,hu2020cooperative}), human-robot interaction (e.g., \cite{van2020adaptive}), and collective transportation and manufacturing (e.g., \cite{du2019leader,fu2022leader}). 

As a prevalent approach, game theory has been widely adopted for studying multi-agent interactions and cooperative control (see \cite{marden2009cooperative}). Most game-theoretic literature leverages Nash games to develop cooperation plans, where all agents are homogeneous in interactions and decision-making (e.g., \cite{mylvaganam2017differential,yao2020self,zhu2020decentralized}). However, it fails to capture the discussed asymmetric structure in guided cooperation. 
Stackelberg games (e.g., \cite{bacsar1998dynamic}) provide a suitable framework for quantifying the heterogeneous capabilities and leader-follower type of interactions in guided cooperation. The corresponding Stackelberg equilibrium solution can be used as an agent-wise optimal control strategy for guided cooperation. Some recent works have investigated asymmetric interactions and cooperation in multi-agent systems based on Stackelberg games. For example, \cite{fisac2019hierarchical} have used feedback Stackelberg games to develop driving strategies for autonomous vehicles to proactively assist the human driver to drive more safely and efficiently. 
\cite{zhao2022stackelberg} have studied the cooperation strategies for guided multi-robot rearrangement tasks based on stochastic Stackelberg games.

Although Stackelberg game-theoretic approaches capture the asymmetric interactions between the prescribed heterogeneous agents (agent-level heterogeneity), it is insufficient to address the following rising control challenges for guided cooperation. First, there is incomplete information about the agents. A leader (agent) may not know the exact model of the follower (agent), demanding learning-based approaches for game-theoretic control. Second, a leader often needs to work with different types of followers for heterogeneous tasks (also known as task-level heterogeneity). As the number of followers increases, designing distinct cooperation plans (using Stackelberg games) for heterogeneous followers demands a tractable and fast adaptive approach.

To address the challenges, we leverage meta-learning to enable learning a customizable plan from a prescribed set of tasks and fast adaptation to a new task with a small amount of learning data (see \cite{hospedales2021meta}).
Meta-learning has been used in many areas to seek adaptive cooperation plans, such as multi-agent systems by \cite{jia2022crmrl}, Internet of Things by \cite{yue2022efficient}, and human robot-interaction by \cite{gao2019fast}. Some recent works have also focused on meta-learning-based control. For example, \cite{harrison2018control} have developed a meta-learning-based approach to stabilize the unseen dynamical system. \cite{richards2021adaptive} have adopted meta-learning to learn a control policy and adapt to unknown environment noise for UAVs to achieve better trajectory tracking. 
Meta-learning also provides a suitable learning mechanism for Stackelberg game-theoretic approaches to address guided cooperation. The leader can learn a meta-knowledge of cooperative control strategies from experience. When a new cooperation task is initiated, the leader can transfer the meta-knowledge to fit the new task only using a small amount of interactive data.

In this work, we establish a Stackelberg meta-learning framework to enable guided cooperative control in linear systems and evaluate the framework using an application of robot teaming. Specifically, a leader guides different types of followers from different starting positions to reach the target destination and forms a team. The framework captures the guided interactions as a dynamic Stackelberg game and uses associated feedback Stackelberg equilibrium (FSE) as the agent-wise optimal control strategy for cooperation. When guiding heterogeneous followers to target destinations, the leader leverages meta-learning to learn a meta-response model for all foreseeable followers and adapts to a customized model for cooperative control when working with a specific follower. We use numerical experiments to corroborate that the proposed Stackelberg meta-learning framework not only enables promising guided control for diverse followers but also achieves a cost-efficient solution compared with other learning approaches and shows better transferability in the learned cooperation strategy compared with individual learning schemes.

\section{Problem Formulation} \label{sec:formulation}
\subsection{Stackelberg Games for Cooperative Control}

We consider that a leader $L$ (she) cooperates with a follower $F$ (he) to complete a task driven by linear-Gaussian dynamics
\begin{equation}
\label{eq:dynamics}
    x_{t+1} = A x_t + B^L u^L_t + B^F u^F_t + w_t,
\end{equation}
where $x_t \in \R^n$ and $u^L_t \in \R^{r^L}$ represents the system state, and the leader's control input at time $t$, $A \in \R^{n\times n}$ and $B^L \in \R^{n\times r^L}$ are state transition matrix and the leader's control input matrix, respectively. Likewise, $u^F_t \in R^{r^F}$ and $B^F \in \R^{n \times r^F}$ are the follower's control and the control input matrix, respectively. Here, $w_t \in \R^n$ are i.i.d. process noise with Gaussian distribution $\mN(0, \bSigma)$. 

Followers have heterogeneous characteristics, distinguished by their type $\theta \in \Theta$. The leader works with one follower at a time to achieve the control task. We assume that the leader does not know the exact type of the follower except for a type distribution $p$ over $\Theta$, where $p(\theta)$ represents the probability that the leader cooperates with a follower with type $\theta$.

The cooperative interactions are strategic. The leader's goal is to minimize the guidance cost $J^L_\theta$ over time horizon $T$ by finding an optimal control trajectory $\bu^{L*} := \{ u^{L*}_t \}_{t=0}^{T-1}$. The less resourceful followers are assumed to be myopic and only minimize one-step cost $J^{F}_\theta$ after observing the current state $x_t$ and the leader's control $u_t$. This asymmetric interaction in the cooperation can be captured by a dynamic Stackelberg game $\mG_\theta$ as follows:
\begin{align}
    \min_{\bu^L} \quad & J^L_\theta(\bu^L) \notag \\
    & \hspace{2mm} := \Exp\left[ \sum_{t=0}^T x_t^\tp Q^L x_t + {u^L_t}^\tp R^L u^L_t + x_t^\tp Q^L_f x_T \right], \label{eq:leader_cost} \\ 
    \text{s.t.} \quad & x_{t+1} = A x_t + B^L u^L_t + B^F_\theta u^{F*}_{\theta,t}(x_t, u^L_t) + w_t, \label{eq:leader_dyn} \\ 
    & \hspace{46mm} t = 0,\dots, T-1, \notag \\
    & u^{F*}_{\theta,t}(x_t, u^L_t) = \arg\min_{u^F} \quad J^F_\theta(u^F; x_t, u^L_t) \label{eq:follower_cost} \\
    & \hspace{0mm} := \Exp[ x_{t+1}^\tp Q^F_\theta x_{t+1} + {u^F}^\tp R^F_\theta u^F], \ t=0,\dots, T-1. \notag
\end{align}
Here, $Q^L \succeq 0, Q^L_f \succeq 0, R^L \succ 0, Q^F_\theta \succeq 0, R^F_\theta \succ 0$ are the leader and the follower's cost matrices with proper dimensions. $B^F_\theta \in \R^{n \times r^F}$ is the type-specific control input matrix.
The follower's problem is captured by \eqref{eq:follower_cost}, where $x_{t+1}$ in \eqref{eq:follower_cost} represents the follower's one-step prediction and evolves according to the dynamics \eqref{eq:dynamics} after the follower observes $x_t$ and $u^L_t$. 
The leader anticipates the follower's response $u^{F*}_{\theta,t}(x_t, u^L_t)$ and uses it for long-term planning. We adopt the FSE $\langle \bu^{L*}(\bx), \bu^{F*}_\theta(\bx) \rangle$ of the game $\mG_\theta$ as the agent-wise cooperation plan for the leader and the follower.
Note that the equilibrium strategies are the functions of the state. The leader and follower can generate controls based on the observed state.

\subsection{Meta Response and Meta-learning Objectives}
Model-based methods such as dynamic programming can find the FSE of $\mG_\theta$ if the leader knows the follower's exact decision-making model (i.e., $J^F_\theta$). However, this information may not be known to the leader. Therefore, the leader needs learning-based approaches to first learn the follower's behavior model and then compute the FSE to find the cooperation plan.

The follower's behavior model can be estimated in various ways, including neural networks. From \eqref{eq:follower_cost}, we can obtain 
\begin{align}
    u^{F*}_{\theta,t}(x_t, u^L_t) &= \notag \\
     -({B^F_\theta}^\tp & Q^F_\theta B^F_\theta + R^F_\theta)^{-1} {B^F_\theta}^\tp Q^F_\theta (Ax_t + B^L u^L_t)  \label{eq:real_br}
\end{align}
for $t=0,\dots, T-1$. It shows the follower's optimal response has a linear structure in $x_t$ and $u^L_t$ under a quadratic cost. We leverage this linear structure and use a matrix parameter $M \in \R^{r^F \times n}$ to estimate the follower's response with $r_\theta: \R^n \times \R^{r^L} \to \R^{r^F}$, defined by
\begin{equation}
\label{eq:br}
    r_{\theta}(x_t, u^L_t; M) = M (Ax_t + B^L u^L_t).
\end{equation}
We substitute the follower's problem \eqref{eq:follower_cost} with \eqref{eq:br}. Then, $\mG_\theta$ becomes a new Stackelberg game, denoted as $\tilde{\mG}_\theta(M)$, where the leader does not know the follower's model but assumes that the follower uses \eqref{eq:br} as his response parameterized $M$. 
Then, the leader uses the FSE $\langle \tilde{\bu}^{L^*}(M), \br^*_\theta(M) \rangle$ of $\tilde{\mG}_\theta(M)$, where $\br^*_\theta(M) := \{ r_{\theta}(x_t, \tilde{u}^{L*}_t; M) \}_{t=0}^{T-1}$, to approximate the one of $\mG_\theta$.
The leader's optimal guidance cost in $\tilde{\mG}_\theta(M)$ is denoted as $\tilde{J}^{L*}(M)$.

The leader faces different game problems when cooperating with heterogeneous followers. It can be cost-prohibitive for the leader to estimate each follower's response model and compute FSEs. Meta-learning provides a learning mechanism to learn from a set of followers and fast adapt to a specific individual follower to achieve cooperative control. Specifically, the leader learns a meta-response from encountered followers as the meta-knowledge of followers' behavior. When working with a new follower, the leader only needs small learning data to adapt the meta-response to the follower-specific one and uses the adapted response model to compute cooperative strategies, i.e., the FSE.

With a slight abuse of notation, we use $M$ as the meta parameter and use \eqref{eq:br} as the meta-response model for all $\theta \in \Theta$. 
We refer to the guided cooperation between the leader and the follower with type $\theta$ as task $\mT_\theta$ to align with the meta-learning context.
A meta-response model should approximate the follower's real behavior (optimal response) and reduce the leader's guidance cost. The latter objective can be quantified by the leader's optimal cost function $\tilde{J}^{L*}(M)$. The former can be achieved by minimizing the response data fitting cost. Let $\mD_\theta = \{ \hat{x}_i, \hat{u}^L_i, \hat{u}^{F*}_i \}_{i=1}^D$ be a best-response data set of $D$ samples collected from the follower with type $\theta$. The data fitting cost is given by
\begin{equation*}
    Q_\theta(M) = \frac{1}{N} \sum_{i=1}^N \norm{M(A \hat{x}_i + B^L \hat{u}^L_i) - \hat{u}^{F*}_i}^2_2.
\end{equation*}
We define the meta-learning objective for the task $\mT_\theta$ as 
\begin{equation}
\label{eq:task_cost}
    L_\theta(M) = \tilde{J}^{L*}_\theta(M) + \gamma Q_\theta(M),
\end{equation}
where $\gamma > 0$ is the weighting parameter.

\subsubsection{Interpretation on $\gamma$}
The weighting parameter $\gamma$ represents how the leader values the follower's response data in learning the follower's behavior model.
When $\gamma = 0$, the leader shows zero interest in the follower's real behavior. She only seeks a unilaterally optimal model, which helps minimize her guidance cost. The learned response model can differ significantly from the follower's real response. When $\gamma \to \infty$, the leader aims to learn the follower's real response as precisely as possible. The approximation accuracy becomes the exclusive objective in meta-learning. Therefore, $\gamma$ provides flexibility in balancing different meta-learning criteria.

\subsection{Meta-Learning as Bilevel Optimization Problems}
The leader uses meta-learning to gain meta-knowledge of followers' behavior and trains an adapted response model for the new coming follower for cooperation.
We split the data $\mD_\theta = \mD^{train}_\theta \cup \mD^{test}_\theta$ following typical learning settings and formulate the meta-learning problem as a bilevel optimization problem (see \cite{rajeswaran2019meta}):
\begin{equation}
\label{eq:ml_out}
    \min_M \quad \Exp_{\theta \sim p}[L_\theta(Z^*_\theta(M); \mD^{test}_\theta)]
\end{equation}
with
\begin{equation}
\label{eq:ml_in}
    Z^*_\theta(M) = \arg\min_Z L_\theta(Z; \mD^{train}_\theta) + \lambda \norm{Z-M}^2_F,
\end{equation}
where $\lambda > 0$ is the weighting parameter. 
The inner-level problem \eqref{eq:ml_in} learns a task-specific optimizer on the training data $\mD^{train}_\theta$. The outer-level problem \eqref{eq:ml_out} improves the generalized performance of the meta parameter on all sampled tasks with data $\mD^{test}_\theta$.

\section{Stackelberg Meta-Learning} \label{sec:sg_meta}

\subsection{Parametric Optimal Control}
We take the linear meta-response \eqref{eq:br} into the dynamics \eqref{eq:leader_dyn}. Then the Stackelberg game $\tilde{\mG}_\theta(M)$ becomes a single-agent linear-quadratic-Gaussian (LQG) control problem:
\begin{equation}
\label{eq:lqg}
\begin{split}
    \min_{\bu^L} \quad & \tilde{J}^L(\bu^L) = \Exp \left[ \sum_{t=0}^T x_t^\tp Q^L x_t + {u^L_t}^\tp R^L u^L_t + x_t^\tp Q^L_f x_T \right], \\ 
    \text{s.t.} \quad & x_{t+1} = \tilde{A} x_t + \tilde{B}^L u^L_t + w_t, \quad, t=0,\dots, T-1,
\end{split}
\end{equation}
where $\tilde{A} := A + B^F_\theta MA$ and $\tilde{B}^L := B^L + B^F_\theta M B^L$.
Given a meta parameter $M$, we can evaluate the leader's optimal guidance cost $\tilde{J}^{L*}(M)$ and the feedback control law $\tilde{\bu}^{L*}(M)$ by solving the discrete Riccati equation
\begin{equation}
\label{eq:riccati}
\begin{split}
    P_{t} = Q^L +& \tilde{A}^\tp P_{t+1} \tilde{A} \\
    -& \tilde{A}^\tp P_{t+1} \tilde{B} (R^L + \tilde{B}^\tp P_{t+1} \tilde{B})^{-1} \tilde{B}^\tp P_{t+1} \tilde{A}
\end{split}
\end{equation}
for $t=0,\dots, T-1$ with $P_T = Q^L_f$. The feedback control $u^{L*}_t = -K_t x_t$ where $K_t := (R^L + \tilde{B}^\tp P_{t+1} \tilde{B})^{-1} \tilde{B}^\tp P_{t+1} \tilde{A}$. The optimal guidance cost $\tilde{J}^{L*}(M) = x_0^\tp P_0 x_0 + \res_0$, where $\res_t = \sum_{j=t+1}^T \tr(\bSigma P_{t+1})$ and $\res_T = 0$.

\subsection{Meta-Response Training}
Solving the inner-level problem \eqref{eq:ml_in} requires optimizing the parameter $M$ over the parameterized cost $\tilde{J}^{L*}(M)$. We have the following proposition to characterize the property of $\tilde{J}^{L*}(M)$.

\begin{prop} \label{prop:1}
With the parametrization of $\tilde{A}$ and $\tilde{B}$ in \eqref{eq:lqg}, the parameterized cost $\tilde{J}^{L*}(M)$ is a rational polynomial of entries of $M$.
\end{prop}

\begin{pf}
See Appendix A.
\end{pf}

Therefore, $\tilde{J}^{L*}(M)$ is continuously differentiable in the entries of $M$, and we can develop gradient methods to solve the inner-level problem. 
To evaluate $\frac{\partial \tilde{J}^{L*}}{\partial M}$, we note that the matrix $P_t$, $t=0,\dots, T-1$, is also parameterized by $M$. Thus, we leverage the Riccati equation \eqref{eq:riccati} to evaluate $\frac{\partial P_t}{\partial M}$ backward from $t=T-1, \dots 0$ with $\frac{\partial P_T}{\partial M} = 0$. $\frac{\partial \res_t}{\partial M}$ can be evaluated similarly.
The convergence of gradient methods on the inner-level problem \eqref{eq:ml_in} is guaranteed because the objective is continuously differentiable in $M$ and is lower bounded by 0 (see \cite{bertsekas1997nonlinear}). The weight $\lambda$ can be used to convexify the inner-level problem and help search for local minimizers.

We use empirical value to approximate the expectation in the outer-level problem \eqref{eq:ml_out} and obtain
\begin{equation}
\label{eq:ml_out_empirical}
    \min_M \quad \frac{1}{\abs{\mT_{batch}}} \sum_{\theta \sim p} L_\theta(Z^*(M); \mD^{test}_\theta).
\end{equation}
Here, $\theta \sim p$ represents the empirical task distribution of sampled batch tasks $\mT_{batch} := \{ \mT_\theta \}$ from $p$.
Following similar computations, we use gradient methods to solve the outer-level problem and find a meta-response model. The iteration follows
\begin{equation}
\label{eq:ml_update}
    M_{k+1} \gets M_k - \frac{\beta}{\abs{\mT_{batch}}} \sum_{\mT_\theta \in \mT_{batch}} \frac{\partial}{\partial M} L_\theta(Z^*_\theta; \mD^{test}_\theta),
\end{equation}
where $\beta > 0$ is the meta-learning step. We summarize the Stackelberg meta-learning algorithm for cooperative control in Alg.~\ref{alg:1}, which outputs a meta-response model.

\begin{algorithm}
\caption{Stackelberg Meta-learning algorithm.}
\label{alg:1}
\begin{algorithmic}[1]
\Require Step $\alpha, \beta$; weight $\gamma, \lambda$; type distribution $p(\theta)$;
\Require Initial mete parameter $M_0$, initial state $x_0$;
\State $k \gets 0$;
\While{$k < \mathrm{MAX\_ITER}$}
    \State Sample a batch of tasks $\mT_{batch} := \{ \mT_\theta \} \sim p$;
    \LineComment{Inner-level problem gradient evaluation}
    \ForAll{task $\mT_\theta \in \mT_{batch}$}
        \State iter $\gets 0$; $Z_\theta \gets M_k$;
        \While{True}
            \State $\frac{\partial P_t}{\partial M}, \frac{\partial \res_t}{\partial M} \gets$ based on \eqref{eq:riccati} and $Z_\theta$ $\forall t$;
            \State $\frac{\partial \tilde{J}^{L*}}{\partial M} \big\vert_{Z_\theta} \gets \frac{\partial}{\partial M}x_0^\tp P_0 x_0 + \frac{\partial \res_0}{\partial M}$;
            \State $\tilde{\bx}(Z_\theta), \tilde{\bu}^{L*}(Z_\theta) \gets$ simulate trajectory;
            \State Randomly sample $N_1$ data;
            \State Sample $N_2$ data around $\tilde{\bx}(Z_\theta), \tilde{\bu}^{L*}(Z_\theta)$;
            \State $\mD^{train}_\theta \gets$ all samples with $N = N_1 + N_2$;
            \State $g \gets \frac{\partial}{\partial M} L_\theta(Z_\theta; \mD^{train}_\theta) + 2\lambda (Z_\theta-M_k)$;
            \State $Z_\theta \gets Z_\theta - \alpha g$;
            \If {iter $> \mathrm{MAX\_GD}$ or $\norm{g} < \epsilon$}
                \State $Z^*_\theta \gets Z_\theta$; break;
            \EndIf
            \State iter $\gets$ iter $+1$;
            \EndWhile
    \EndFor
    \LineComment{Outer-level problem gradient evaluation}
    \ForAll{task $\mT_\theta \in \mT_{batch}$}
        \State $\frac{\partial P_t}{\partial M}, \frac{\partial \res_t}{\partial M} \gets$ based on \eqref{eq:riccati} and $Z^*_\theta$ $\forall t$;
        \State Sample $\mD^{test}_\theta$ (same sampling rule as $\mD^{train}_\theta$);
        \State Compute $\frac{\partial}{\partial M} L_\theta(Z^*_\theta; \mD^{test}_\theta)$;
    \EndFor
    \State Update $M_{k+1}$ by \eqref{eq:ml_update};
    \State $k \gets k + 1$;
\EndWhile
\\
\Return $M_{meta} \gets M_k$;
\end{algorithmic}
\end{algorithm}

\subsubsection{Sampling Follower's Response Data}
We note that the inner-level problem \eqref{eq:ml_in} shows that the meta parameter is updated within a small neighborhood of the original one due to the regularization term. Hence, the updated leader's trajectory will likely stay near the previous one. The samples near the trajectory can better help the leader refine the follower's response model near the trajectory and hence make a better update. This sampling technique is more useful when the leader uses a nonlinear response mode such as neural networks to estimate the follower's behavior.
We set $\kappa := N_2 / N_1$ to control the sample ratio in $\mD^{train}_\theta$ and $\mD^{test}_\theta$.

\subsection{Response Adaptation}
Using the meta parameter $M_{meta}$ and the meta-response model from Alg.~\ref{alg:1}, the leader can fast adapt to a new coming follower using a small amount of data samples. Specifically, the leader samples a type-specific data set $\mD'_\theta$ using $M_{meta}$ when she starts cooperating with a follower with type $\theta$.
Then she customizes a response parameter $M^*_\theta$ from $M_{meta}$ to adapt to the follower by solving
\begin{equation}
\label{eq:ml_adapt}
    M^*_\theta = \arg\min_M L_\theta(M_{meta}; \mD_\theta') + \eta \norm{M - M_{meta}}^2_F,
\end{equation}
where $\eta \geq 0$ is the regularization weight. In practice, we can select $\eta = \lambda$.

\section{Experiments and Evaluations} \label{sec:experiment}

In this section, we demonstrate our Stackelberg meta-learning framework using a case study in cooperative robot teaming, where a leader robot guides the follower robot to a target destination to form a team. Let $x^L = [p^L, v^L] \in \R^4, u^L \in \R^2$ ($x^F = [p^F, v^F] \in \R^4$, $u^F \in \R^2$) be the leader's (the follower's) position, velocity, and control input. The joint state $x := [x^L, x^F]$. We assume the leader and the follower have a double integrator dynamics, where $\ddot{p}^L = u^L$ and $\ddot{p}^F = u^F$. The corresponding discrete dynamical systems are obtained by setting a discretization time $\dd t=0.5$. We set the control time horizon $T=10$ and the target destination $x^\dd = 0$. $w_t \in \R^8$ are i.i.d. Gaussian noise $\sim \mN(0, 0.5I)$. We consider five types of followers with a type distribution $p = [0.2, \ 0.3, \ 0.1, \ 0.2, \ 0.2]$.

\subsection{Meta-learning Results}
\begin{figure}[h]
    \centering
    \begin{subfigure}[b]{0.24\textwidth}
        \centering
        \includegraphics[height=3.2cm]{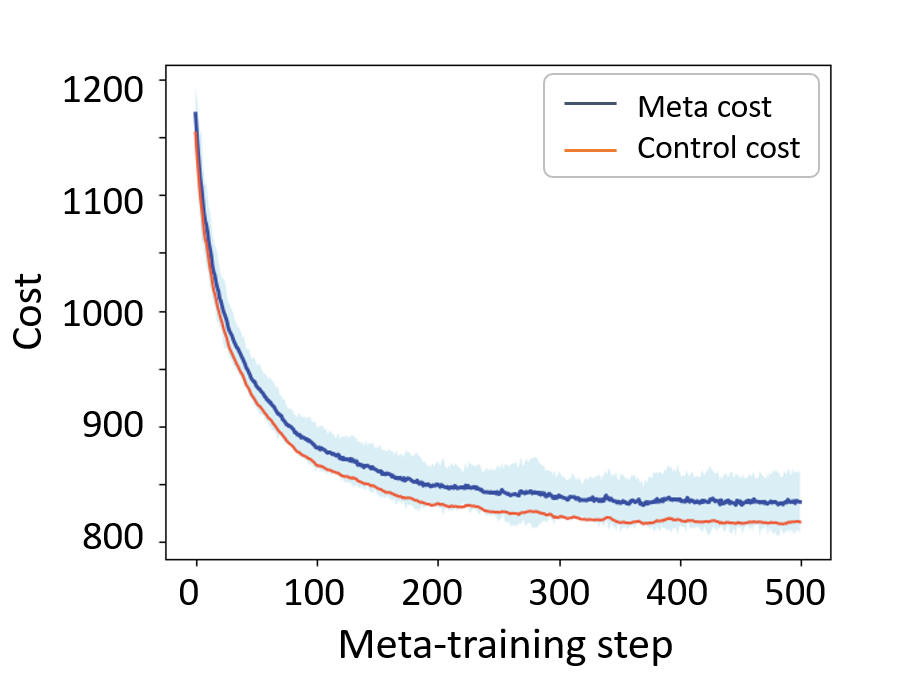}
        \caption{Meta-training process.}
        \label{fig:meta.1}
    \end{subfigure}
    \begin{subfigure}[b]{0.24\textwidth}
        \centering
        \includegraphics[height=3.2cm]{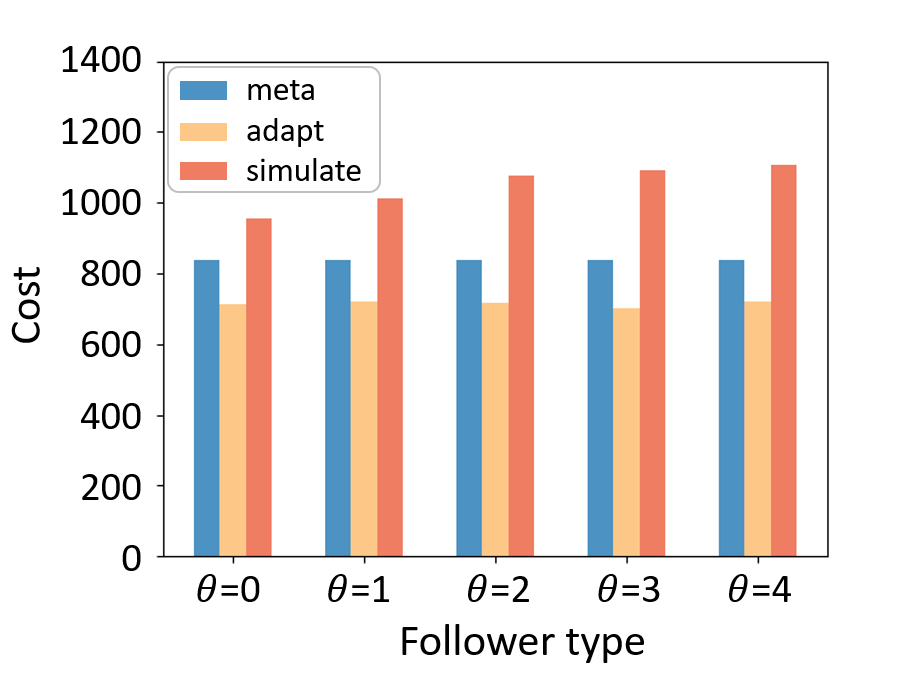}
        \caption{Adaptation comparison.}
        \label{fig:meta.2}
    \end{subfigure}
    \caption{Meta training and adaptation results.}
    \label{fig:meta}
\end{figure}

We set $\kappa = 2$ and use $N=6$ response data in each iteration to perform meta-training. The hyperparameters are set by $\gamma = 5, \lambda = 100$. The training process is evaluated by the empirical meta-cost used in \eqref{eq:ml_out_empirical}. We conduct $20$ simulations with a randomly generated initial guess $M_0$ and plot the mean-variance training result in Fig.~\ref{fig:meta}. 

Fig.~\ref{fig:meta.1} shows that the meta-learning algorithm reduces the meta-cost and converges to a local minimum. The mean value of the leader's optimal guidance cost $\tilde{J}^{L*}$ (orange line) is also reduced as the meta-training proceeds, which means that the meta-response model becomes more efficient for the leader to perform the guidance. The variance comes from different sampled response data in each simulation to train the meta model.

The adapted results for different types of followers are shown in Fig.~\ref{fig:meta.2}. The blue bar represents the leader's expected guidance cost $\tilde{J}^{L*}$ using the meta-response model before the adaptation, serving as a baseline. The yellow bar shows the expected guidance cost $\tilde{J}^{L*}$ using the adapted response model for different types of followers, respectively. As expected, the adapted model provides a lower guidance cost for the leader than the baseline.

\begin{figure}
    \centering
    \begin{subfigure}[b]{0.24\textwidth}
        \centering
        \includegraphics[height=3.2cm]{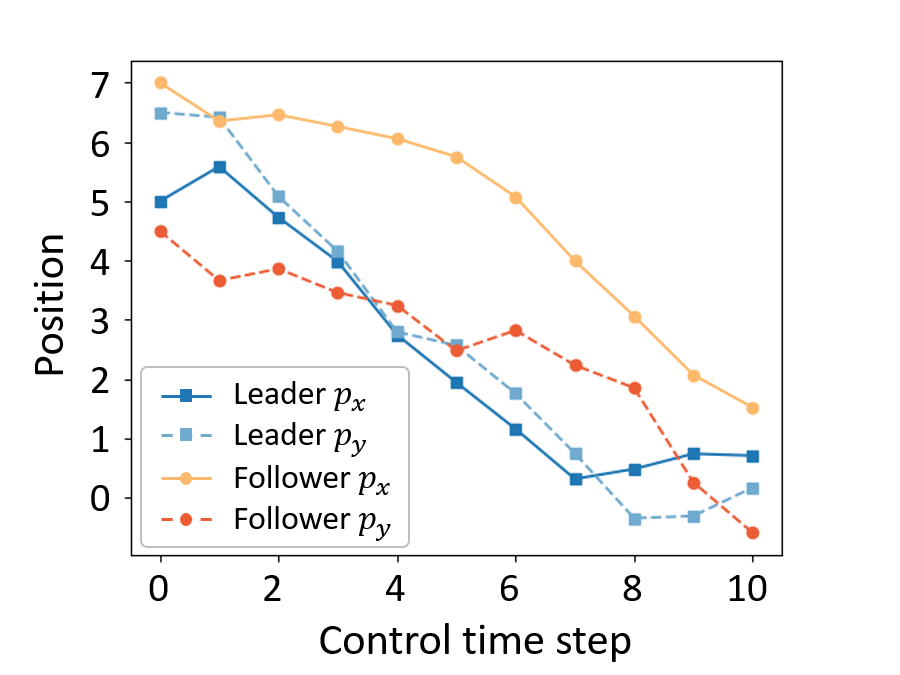}
        \caption{Position trajectories.}
        \label{fig:traj.1}
    \end{subfigure}
    \begin{subfigure}[b]{0.24\textwidth}
        \centering
        \includegraphics[height=3.2cm]{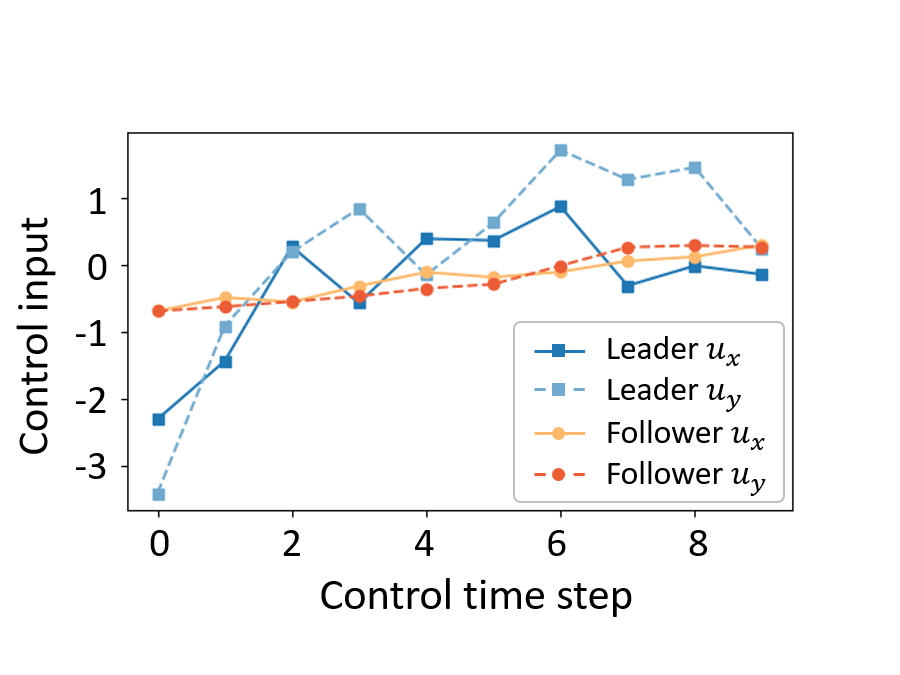}
        \caption{Control trajectories.}
        \label{fig:traj.2}
    \end{subfigure}
    \caption{Trajectories for $\theta=0$ follower after adaptation.}
    \label{fig:traj}
\end{figure}

Due to the process noise and the estimation error, the follower's real behavior can deviate from the leader's expectation.
We simulate interactive trajectories to view the real cooperation performance, where the leader uses the adapted response model to design control strategies, and the follower uses his true model \eqref{eq:follower_cost} to respond. The leader's simulated guidance costs for each type of follower are shown by the red bar in Fig.~\ref{fig:meta.2}.
We observe that the simulated costs are higher but not significantly greater than the expected costs. It shows that the adapted response model and the resulting control strategies can provide satisfactory results in guidance tasks. 
For simplicity, we plot the position and control trajectories for the leader and the follower with type $\theta=0$ in Fig.~\ref{fig:traj}. The leader and the follower start from $[5, 6.5]$ and $[7, 4.5]$, respectively.
From Fig.~\ref{fig:traj.1}, the leader can design effective cooperative strategies using the adapted model to guide the follower to approach the zero state. Their controls also approach 0 by the end of the guidance. The trajectory convergence direction shows that the guidance is effective.

\subsection{Comparison with Unilateral Learning}
The unilateral-learning approach refers to the leader learning a response model based solely on the guidance cost instead of the follower's real response. It is equivalent to set $\gamma = 0$ in \eqref{eq:task_cost}. 
Since followers' response is not involved, the learned response models are the same for all followers. Therefore, the learning is fast and has an average training time of less than 1 min, which contrasts sharply with the meta-learning approach, with an average time of 32 min.

\begin{figure}
    \centering
    \begin{subfigure}[b]{0.24\textwidth}
        \centering
        \includegraphics[height=3.2cm]{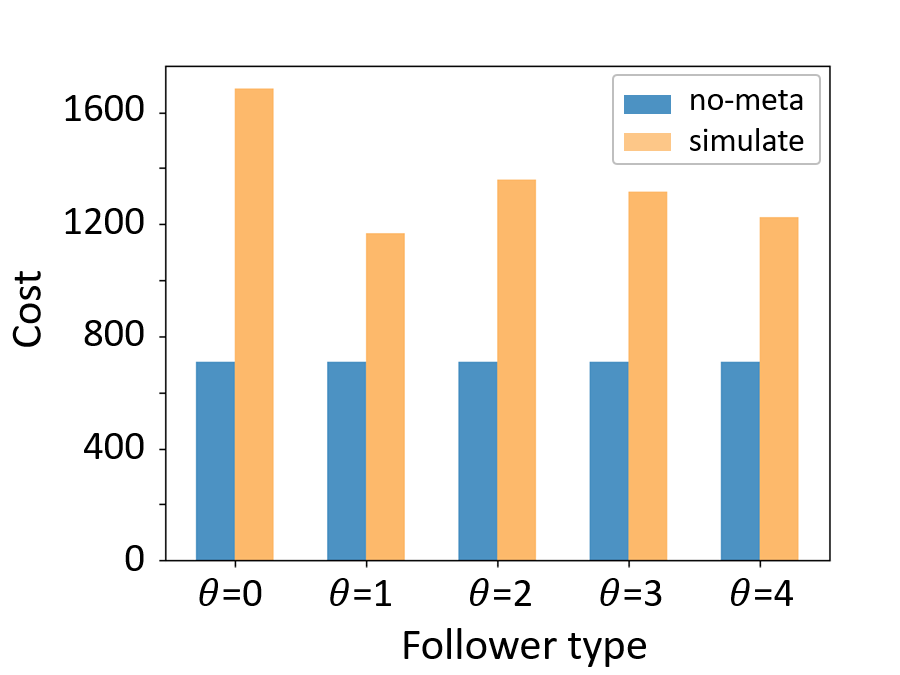}
        \caption{Leader's guidance cost using unilateral learning.}
        \label{fig:nometa.1}
    \end{subfigure}
    \begin{subfigure}[b]{0.24\textwidth}
        \centering
        \includegraphics[height=3.2cm]{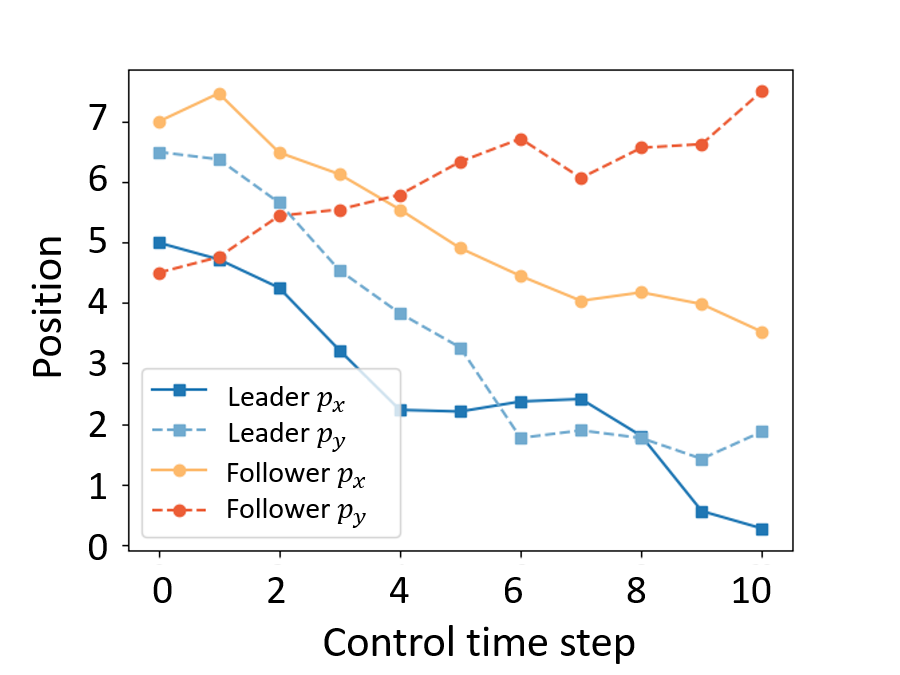}
        \caption{Trajectories using unilateral learning for type $\theta=2$ follower.}
        \label{fig:nometa.2}
    \end{subfigure}
    \caption{Costs and trajectories for unilateral learning.}
    \label{fig:nometa}
\end{figure}

We evaluate the leader's expected (blue) and simulated (yellow) guidance cost using the model obtained from the unilateral-learning approach and compare them in Fig.~\ref{fig:nometa.1}. 
The simulated cost significantly deviates from the expected one, indicating that the learned model is less effective in the guidance task. Besides, the expected and simulated costs are also greater than the counterparts in meta-learning (see Fig.~\ref{fig:meta.2}), showing the adapted models outperform the unilaterally learned ones. 
For simplicity, we show the simulated trajectory for the leader and the follower with type $\theta=2$ in Fig~\ref{fig:nometa.2}. The unilateral-learning approach fails to guide the follower to the origin. Instead, the follower moves in the opposite direction, resulting in a failure in the guidance task.
Although unilateral learning saves considerable training time compared with the meta-learning approach, it can significantly sacrifice the model accuracy and the guidance performance.

\subsection{Individual Learning and Transferability}
\begin{figure}
    \centering
    \begin{subfigure}[t]{0.24\textwidth}
        \centering
        \includegraphics[height=3.2cm]{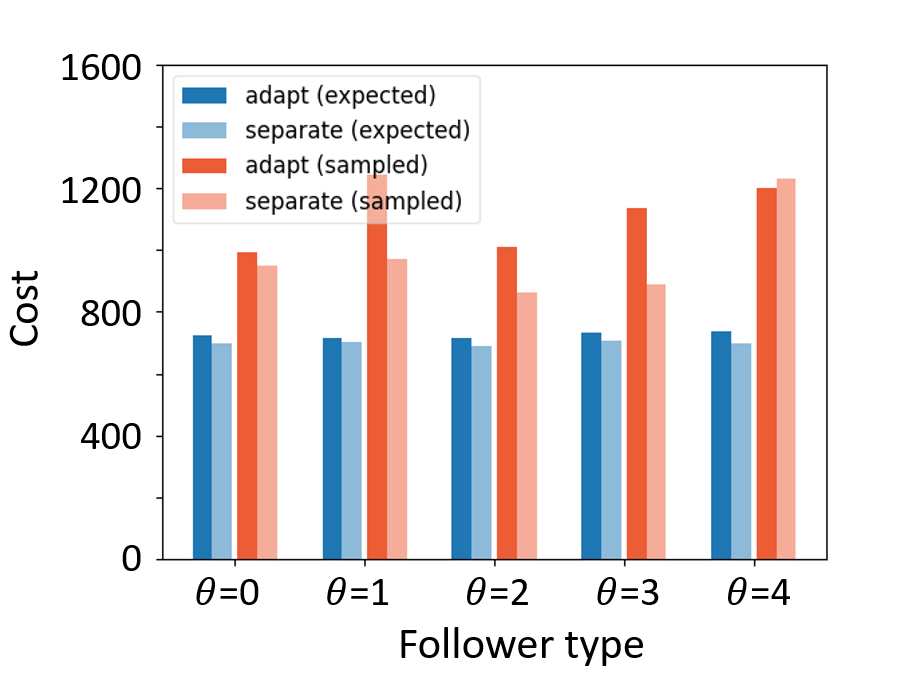}
        \caption{Comparison with meta-learning.}
        \label{fig:trans.1}
    \end{subfigure}
    \begin{subfigure}[t]{0.24\textwidth}
        \centering
        \includegraphics[height=3.2cm]{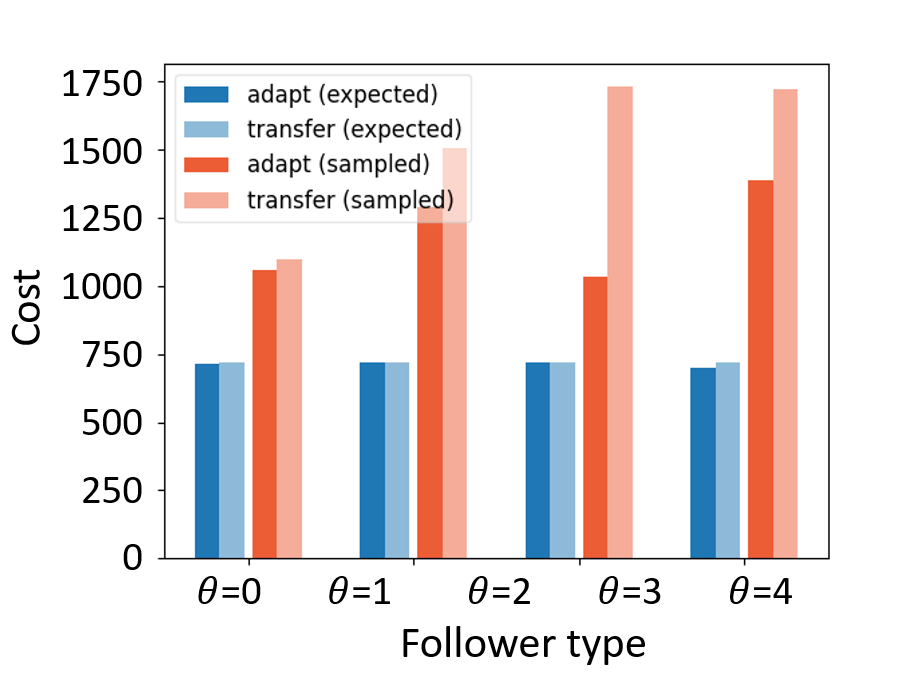}
        \caption{Adapting $\theta=4$ follower's model to the rest.}
        \label{fig:trans.2}
    \end{subfigure}
    \caption{Results for individual learning.}
    \label{fig:trans}
\end{figure}

The individual-learning approach refers to the leader learning separate response models for every follower and generating different guidance strategies.
We evaluate the leader's expected (blue bars) and simulated (red bars) guidance cost in Fig.~\ref{fig:trans.1}. We also plot our meta-learning result with dark colors for comparison. 
It is not surprising that individual learning provides slightly smaller guidance costs compared with meta-learning because it trains designed models for different followers. 

However, meta-learning can adapt the meta-response model to a specific follower and provide good guidance. Individual learning does not have such flexibility and transferability. 
To see this, we adapt the learned model of the follower $\theta=4$ obtained by the individual-learning approach to other followers by following adaptation rule \eqref{eq:ml_adapt}. We evaluate the leader's expected and simulated guidance cost in Fig.~\ref{fig:trans.2}. 
We can observe that meta-learning provides smaller expected and simulated guidance costs, showing that adapted models are more efficient for the leader in performing guidance control. 
The smaller simulated costs indicate that the meta-learning approach outperforms individual learning in real guidance tasks.
Besides, individual learning requires considerable learning resources, especially when there are many followers. The meta-learning approach can leverage its better transferability in the learned response model, providing a more flexible and faster adaptation for different guidance tasks. 

\section{Conclusion} \label{sec:conclusion}
We have proposed a Stackelberg meta-learning framework for guided cooperative control in LQG systems. Our framework not only captures the leader-follower type of interactions in guided cooperation but also provides a learning mechanism to adapt to different guided control tasks. The case study in robot teaming application has demonstrated that the framework successfully provides effective and transferable guidance control strategies to accomplish different guidance tasks.
As we have observed in the simulation, although a learned cooperation strategy can guide the follower toward the destination, it cannot perform as precisely as deterministic control. How to guarantee the control performance within an allowable range would be a valuable future research direction.
For other future work, we would generalize our framework to more general control systems and investigate analytic properties such as optimality conditions and sample complexity.

{
\bibliography{mybib}
}

\newpage
\appendix

\section{Proof of Proposition \ref{prop:1}} \label{app:prop1_prof}

We denote $\poly(n, M)$ as the set of polynomials that uses the entries of $M$ as arguments and has the highest order $n$. For example, $m_{11}^3 m_{12}^2 + 2m_{21}^3 - m_{22}^4 +1 \in \poly(5, M)$. We further use a generalized fraction $\frac{\poly(n, M)}{\poly(m, M)}$ to denote the class of rational polynomial whose numerator belongs to $\poly(n, M)$ and denominator belongs to $\poly(m, M)$. For example, $\frac{m_{11}^2 m_{22}^2 + m_{32}m_{43}+1}{m_{23}^2 m_{31} + 2} \in \frac{\poly(4, M)}{\poly(3, M)}$. Any polynomial in $\poly(n,M)$ has the same highest-order $n$ regardless of the combination of arguments. For example, $m_{12} m_{22}$ and $m_{31}^2+m_{32}$ are both in $\poly(2, M)$. For simplicity, we write $\poly(n) := \poly(n, M)$. We write $A_{ij}$ as the $ij$-entry of $A$. 

We first introduce the following lemma.
\begin{lem}[Lax (2007)] \label{lemma:1}
For any invertible matrix $A \in \R^{n\times n}$, we have $A^{-1}_{ji} = (-1)^{i+j} \frac{\det([C_{ij}])}{\det(A)}$, where $[C_{ij}]$ is the submatrix obtained by deleting $i$-th row and $j$-th column of $A$. The determinant of $A$ can be computed by the Leibniz formula 
\begin{equation}
\label{eq:determinant}
    \det(A)=\sum _{\tau \in S_{n}} \sgn(\tau ) \prod_{i=1}^{n}a_{i,\,\tau (i)},
\end{equation}
where $S_n$ is the set of all permutation of the set $\{1,2,\dots ,n\}$ and $\sgn(\tau)$ is the sign function that returns either $+$ or $-$ for each permutation $\tau \in S_n$.
\end{lem}

From the parameterization of $\tilde{A}$, we observer that the entry $\tilde{A}_{ij} \in \poly(1)$ for all $i,j$, i.e., $\tilde{a}_{ij} = c_0 + \sum_{ij} c_{ij} m_{ij}$ for some constant $c_0$ and $c_{ij}$. Likewise, $\tilde{B}_{ij} \in \poly(1)$ for all $i,j$.
Let $X_{t} := R^L + \tilde{B}^\tp P_{t} \tilde{B} \in \R^{r^L \times r^L}$ for $t=1,\dots, T$. It is clear that $X_t \succ 0$ and hence $\det(X_t) > 0$, $t=1,\dots, T$.

When $t = T$, we have $P_T = Q^L_f$ which is constant. Since the multiplication between two polynomials produces another polynomial whose highest order is the sum of the highest order of two multiplicands, we have $(X_T)_{ij} \in \poly(2)$ for all $i,j$. From Lemma \ref{lemma:1}, we have $\det(X_T) \in \poly(2r^L)$. Let $[C_{ij}]$ be the submatrix obtained from $X_T$ by deleting the $i$-th row and the $j$-th column. Using \eqref{eq:determinant}, we have $\det([C_{ij}]) \in \poly(2(r^L-1))$. Therefore, the entry $(X^{-1}_T)_{ij} \in \frac{\poly(2(r^L-1))}{\poly(2r^L)}$ for all $i,j$. In other words, the entry of $X^{-1}_T$ is a rational function of entries of $M$. Note that every entry $(X^{-1}_T)_{ij}$ has the same denominator $d_T := \det(X_T) > 0$. 
Since every entry of $\tilde{A}, \tilde{B}$ and $X_T$ belongs to the same polynomial class, respectively, we can conclude from \eqref{eq:riccati} that $(P_{T-1})_{ij}$ is also a polynomial and $(P_{T-1})_{ij} \in \frac{\poly(2(r^L+1))}{\poly(2r^L)}$ for all $i,j$. This can be obtained by performing matrix multiplication. Besides, all entries $(P_{T-1})_{ij}$ have the same denominator $d_T$.

When $t=T-1$, it is clear that $(X_{T-1})_{ij} \in \frac{\poly(2r^L+4)}{\poly(2r^L)}$ for all $i,j$. Then we have $\det(X_{T-1}) \in \frac{\poly((2r^L+4)r^L)}{\poly(2r^L \cdot r^L)}$. Let $[D_{ij}]$ be the submatrix obtained from $X_{T-1}$ by deleting the $i$-th row and $j$-th column. $\det([D_{ij}]) \in \frac{\poly((2r^L+4)(r^L-1))}{\poly(2r^L \cdot (r^L-1))}$. Note that $\det(X_{T-1})$ and $\det([D_{ij}])$ have a common divisor $(d_T)^{r^L-1}$ in the denominator. By canceling the common divisor, we obtain $(X^{-1}_{T-1})_{ij} \in \frac{\poly((2r^L+4)(r^L-1)+r^L)}{\poly((2r^L+4)r^L)}$. Besides, every entry $(X^{-1}_{T-1})_{ij}$ has the same denominator $d_{T-1} \in \poly((2r^L+4)r^L)$ and $d_{T-1} > 0$. 
Since every entry of $P_{T-1}$ and $X_{T-1}$ belongs to the same polynomial class, respectively, we can conclude from \eqref{eq:riccati} that $(P_{T-2})_{ij}$ is also a polynomial and $(P_{T-2})_{ij} \in \frac{\poly((2r^L+4)(r^L+1))}{\poly((2r^L+4)r^L)}$ for all $i,j$. Every entry $(P_{T-1})_{ij}$ has the the same denominator $d_{T-1}$.

By induction, we have 
\begin{equation*}
    (P_0)_{ij} \in \frac{\poly(\sum_{t=0}^T 2(r^L+1)^t)}{\poly(\sum_{t=0}^T 2(r^L+1)^t-2T)}, \quad \forall i,j,
\end{equation*}
The denominator of $J^{L*}(M)$, which is the denominator of $(P_0)_{ij} \ \forall i,j$, is always positive. This completes the proof.

\section{Evaluating Matrix Derivatives} \label{app:matrix_derivative}
This appendix discusses numerical details on computing matrix derivatives used in  Alg.~\ref{alg:1}.

Let $f: \R^{p\times q} \to \R^{m\times n}$ is a differentiable function, i.e., each element $f_{ij}(X), i=1,\dots, m, j=1,\dots,n$ is differentiable in its argument.
For simplicity, we denote $i=1,\dots, m$ as $i \in \{m\}$. Similarly, $i,j \in \{m,n\}$ represents $i=1,\dots, m, j=1,\dots,n$. We use $D_Xf$ to represent the derivatives $\frac{\partial f(X)}{\partial X}$.

\subsection{Matrix Derivative Layout} \label{app:1.layout}
The derivative of $f$ can be computed and referenced by a scalar derivative $D_X f_{ij,kl} = \frac{\partial f_{ij}}{\partial X_{kl}}$, $i,j \in \{m,n\}$, $k,l \in \{p,q\}$. The problem is how to design the layout of $D_X f$. 
One direct layout is 
\begin{equation}
\label{eq:df_layout}
    D_X f = \begin{bmatrix} 
        \frac{\partial f_{11}}{\partial X} & \cdots & \frac{\partial f_{1n}}{\partial X} \\ 
        \vdots & \ddots & \vdots \\
        \frac{\partial f_{m1}}{\partial X} & \cdots & \frac{\partial f_{mn}}{\partial X}
    \end{bmatrix},
\end{equation}
where $\frac{\partial f_{ij}}{\partial X}$ is a $p\times q$ matrix having the same size as $X$ and its $(k,l)$-element is $\frac{\partial f_{ij}}{\partial X_{kl}}$. We use $D_X f_{ij,:}$ to denote the $(i,j)$-block matrix $\frac{\partial f_{ij}}{\partial X} \in \R^{p\times q}$.
Different layouts exist for matrix derivatives (e.g., Vetter (1970); Magnus and Neudecker (1985)), 
and Magnus-Neudecker (M-N) convention is commonly used. The M-N convention vectorizes $f$ and $X$ by stacking their columns into a vector. Then the matrix-valued function becomes a vector-valued function $\operatorname{vec}f(\operatorname{vec} X)$, and its derivative is a standard $mn \times pq$ Jacobian matrix. The M-N convention provides advantages for theoretical analysis. 
However, the direct layout \eqref{eq:df_layout} is easier to manage for computation. Here, we should treat $D_X f$ in \eqref{eq:df_layout} as a four-dimensional (4D) tensor instead of a large two-dimensional (2D) matrix because it has four independent index axes. Then we can apply common arithmetic rules on its first two index axes for computation. 

The layout \eqref{eq:df_layout} is also valid to represent any 4D tensors in $\R^{(m\times n) \times (p\times q)}$, whose $(i,j)$-th element is a matrix in $\R^{p\times q}$, $i,j \in \{m,n\}$.

\subsection{Derivative of Matrix Multiplication} \label{app:1.multiply}
We define an operator $\star$ performing multiplication on a 2D matrix and the first two dimensions of a 4D tensor obeying the layout \eqref{eq:df_layout}. Let $W = U \star V$ and $U \in \R^{(m\times r) \times (p\times q)}$ and $V \in \R^{r\times n}$. Then $W_{ij} = \sum_{r=1}^k U_{ir,:} V_{rj}$, $i,j \in \{m,n\}$. Likewise, let $W' = U' \star V'$ and $U' \in \R^{m\times r}$, $V' \in \R^{(r\times n) \times (p\times q)}$. Then $W'_{ij} = \sum_{r=1}^k U_{ir} V_{rj,:}$, $i,j \in \{m,n\}$.

Let $f(X) = Y(X)Z(X)$ with $Y: \R^{p\times q} \to \R^{m \times r}$ and $Z: \R^{p\times q} \to \R^{r \times n}$. Since $f_{ij}(X) = \sum_{k=1}^r Y_{ir}(X) Z_{rj}(X)$, we take the derivative and obtain $D_X f_{ij} = \sum_{k=1}^r (D_X Y_{ir}) Z_{rj} + Y_{ir} (D_X Z_{rj})$,
$i,j \in \{m,n\}$.
We can also compute $D_X Y$ and $D_X Z$ as 4D tensors obeying the layout \eqref{eq:df_layout}. If $D_X f$ follows the same layout, we can write $D_X f = D_X Y \star Z + Y \star D_X Z$.

Therefore, from the Riccati equation \eqref{eq:riccati}, we can compute  
\begin{equation*}
    \frac{\partial \tilde{A}^\tp P_{t+1} \tilde{A}}{\partial M} = \frac{\partial \tilde{A}^\tp}{\partial M} \star (P_{t+1} \tilde{A}) + \tilde{A}^\tp \star \frac{\partial P_{t+1}}{\partial M} \star \tilde{A} + \tilde{A}^\tp P_{t+1} \star \frac{\partial \tilde{A}}{\partial M},
\end{equation*}
where 
\begin{equation*}
    \frac{\partial \tilde{A}}{\partial M} = B^F_\theta \star \frac{\partial M}{\partial M} \star A, \quad  
    \frac{\partial \tilde{B}}{\partial M} = B^F_\theta \star \frac{\partial M}{\partial M} \star B^L.
\end{equation*}
We can verify that $D_M M_{ij, kl} := \frac{\partial M_{ij}}{\partial M_{kl}} = 1$ if $i=k, j=l$ and is 0 otherwise.

\subsection{Derivative of Matrix Inverse} \label{app:1.inverse}
For a square matrix $W$, the matrix identity tells
\begin{equation}
\label{eq:dW_inv}
    \frac{\partial W^{-1}}{\partial x} = - W^{-1} \frac{\partial W}{\partial x} W^{-1},
\end{equation}
where $x \in \R$ is a scalar variable. Now let $W: \R^{m\times n} \to \R^{r \times r}$ and assume $W(X)$ is always invertible. We can evaluate $\frac{\partial W^{-1}(X)}{\partial X_{kl}}$ with \eqref{eq:dW_inv} for $k,l \in \{p,q\}$. 
To use the layout \eqref{eq:df_layout} for $D_X W^{-1}$, we can extract all $(i,j)$-element from $D_X (W^{-1})_{:,kl}$ block matrices for all $k,l\in\{p,q\}$ and form a new $m\times n$ matrix $D_X {W^{-1}}_{ij,:}$. We repeat the process for all $i,j \in \{r,r\}$ to construct $D_X W^{-1}$.

To compute the derivative of $(R + \tilde{B}^\tp P_{t+1} \tilde{B})^{-1}$, 
we first let $W = R + \tilde{B}^\tp P_{t+1} \tilde{B}$ and compute $W^{-1}$ and $D_M W$ with Appendix \ref{app:1.multiply}. Then we evaluate $D_M (W^{-1})_{:,kl}$ for $k,l \in \{r^F,n\}$ and rearrange the result to obtain $D_M W^{-1}$.

\subsection{Complexity} \label{app:1.complexity}
Complexity reveals the relationship between the number of elementary operations of an algorithm and the input data size.
For example, we use the same definition of $Y$ and $Z$ in Appendix \ref{app:1.multiply}. Computing $Y\star D_X Z$ and $D_X Y \star Z$ yield a complexity of $\mO(mn rpq)$. Compared with the normal matrix multiplication $YZ$, which has a complexity of $\mO(mnr)$, the additional order $pq$ comes from the inner matrix multiplication. i.e., calculating the product $(D_X Y_{ij,:}) Z_{ji}$ requires $\mO(pq)$. We further let $W \in \R^{n\times w}$. Then computing $Y \star D_X Z \star W$ has a complexity of $\mO(mnpq(r+w))$. In comparison, computing the matrix product $YZW$ has a complexity of $\mO(mn(r+w))$. 
Let $W \in \R^{r\times r}$. evaluating $W^{-1}$ requires $\mO(r^3)$ operations. From the analysis in Appendix ~\ref{app:1.inverse}, computing $D_X W^{-1}$ requires $\mO(r^3) + \mO(r^3 pq)$ operations.

In the Riccati equation \eqref{eq:riccati}, we note that $D_M M$ has a simple structure.
$D_M M \star A$ simply extracts the columns of $A$ to form inner block matrices with proper order and thus has a complexity of $\mO(1)$. Let $W = D_M M \star A$. $W_{ij,:}$ is a sparse matrix where the $j$-th row equals to the $i$-th column of $A$. The same applies to $B^F_\theta \star D_M M$, which extracts the rows of $B^F_\theta$ to form inner block matrices. Hence, the complexity of computing $D_M \tilde{A}$ can be reduced to $\mO(n^3 r^F)$. Likewise, computing $D_M \tilde{B}$ has a complexity of $\mO(n^2 r^L r^F)$. 
Therefore, using the differentiation rule and the inverse formula, evaluating $D_M P_t$ provides a complexity of $\mO(nr^F [n^3+(r^L)^3 + n^2 r^L + n(r^L)^2] )$. For comparison, computing $P_t$ gives a complexity of $\mO(n^3 + (r^L)^3 + n^2 r^L + n(r^L)^2$.

\vspace{3mm}
\begin{center}
APPENDIX REFERENCES    
\end{center}

Lax, P.D. (2007).
\newblock \emph{Linear Algebra and Its Applications}.
\newblock Wiley, 2 edition.

Vetter, W. (1970).
\newblock Derivative operations on matrices.
\newblock \emph{IEEE Transactions on Automatic Control}, 15(2), 241--244.

Magnus, J.R. and Neudecker, H. (1985).
\newblock Matrix differential calculus with applications to simple, hadamard,
  and kronecker products.
\newblock \emph{Journal of Mathematical Psychology}, 29(4), 474--492.


\end{document}